\documentclass[11pt]{article}
\usepackage{a4wide}
\usepackage{latexsym}
\usepackage{epsfig}
\usepackage{amsmath}
\usepackage{cite}
\usepackage{amssymb}
\usepackage{graphicx}

\long\def\symbolfootnote[#1]#2{\begingroup%
\def\thefootnote{\fnsymbol{footnote}}\footnote[#1]{#2}\endgroup}

\def\as{\alpha_s}

\begin{document}

\thispagestyle{empty}

\begin{flushright}
  NIKHEF/2012-007\\ ITP-UU-12/12\\
IMSc 2012/4/6\\FR-PHENO-2012-006
\end{flushright}

\vspace{1.5cm}

\begin{center}
 {\Large\bf Soft-collinear effects for prompt photon production\\[2ex] 
 via fragmentation}\\[1cm] 
  {\bf  Rahul Basu$^{a}\symbolfootnote[2]{Deceased.}$, Eric Laenen$^{b}$, Anuradha Misra$^c$, Patrick Motylinski$^d$}\\[1cm]
 $^a$ {The Institute of Mathematical Sciences, CIT Campus,
  Taramani, Chennai 600 113, India}\\[0.4cm]
  $^b${ITFA, Science Park 904, 1090 GL Amsterdam, \\
  ITF, Utrecht University, Leuvenlaan 4, 3584 CE Utrecht\\
  Nikhef, Science Park 105, 1098 XG Amsterdam, The
  Netherlands}\\[0.4cm]
 $^c$ {Department of Physics, University of Mumbai, Santacruz(E),
   Mumbai, 400\,098, India}\\[0.4cm]
 $^d$ {Physikalisches Institut, Albert-Ludwigs-Universit\"{a}t Freiburg,
Hermann-Herder-Stra{\ss}e 3, D-79104 Freiburg i.Br., Germany}\\

\end{center}

\vspace{1.5cm}
\begin{center}
  \textit{This paper is dedicated to the memory of Rahul Basu.}
\end{center}
\vspace{1cm}

\begin{abstract}
\noindent{
  We study the impact of leading soft-collinear effects on threshold
  and joint-resummed calculations for the production of prompt photons
  via parton fragmentation, complementing a previous study for
  direct production. We assess these effects for both fixed-target and
  collider kinematics.  We find them to be small, but noticeable and
  comparable to the direct case.}
\end{abstract}

\vspace*{\fill}

\newpage
\reversemarginpar

\section{Introduction}
\label{sec:introduction}

The perturbative QCD description of prompt production at hadron
colliders can involve sizeable corrections from soft and
collinear parton emission.  In particular, the presence of a threshold
at fixed $p_T$ induces large logarithmic corrections
\cite{Sterman:1987aj,Catani:1989ne}. Expressed in terms of a (Melllin)
moment variable $N$ with the property that as $N \rightarrow \infty$ the
kinematics approaches the threshold limit, these corrections
take the form ($L = \ln N$),
\begin{equation}
  \label{eq:29}
  \alpha_s^i \sum_j^{2i}\, a_{ij} L^j\,,
\end{equation}
where the $a_{ij}$ depend in general on the process.  Such large
logarithmic corrections can be organized and controlled through
all-order resummation, for threshold
\cite{Laenen:1998qw,Catani:1998tm,Catani:1999hs,Kidonakis:1999hq,Bolzoni:2005xn,Becher:2009th}
and joint \cite{Laenen:2000de,Laenen:2000ij,Sterman:2004yk,Li:1998is,Sterman:2000pt}
resummation.

In recent years, in the context of threshold
resummation, other large classes of terms have been brought under
all-order control, such as large constants (``$\pi^2$ terms'') 
\cite{Parisi:1980xd,Magnea:1990zb,Eynck:2003fn} and a series of
so-called ``soft-collinear'' terms
\begin{equation}
  \label{eq:30}
  \alpha_s^i \sum_j^{2i-1}\, d_{ij} \frac{\ln^jN}{N}\,.
\end{equation}
In a previous study \cite{Basu:2007nu} we performed the resummation of
the leading terms ($j=2i-1$) for direct production of prompt photons.  In this paper we extend this study to 
the leading soft-collinear terms (which we shall also refer to as
``$\ln N/N$ terms'') in the
complimentary case of prompt photon production by fragmentation. In
this case more subprocesses contribute, and their
color states must be accounted for.  These issues occur as well in
threshold resummation without soft-collinear effects for both mechanisms of prompt photon production
\cite{deFlorian:2005wf}. Subleading soft-collinear terms have been
studied recently in other contexts in Refs.\cite{Laenen:2008ux,Laenen:2008gt,Laenen:2010uz,Gardi:2010rn,Grunberg:2009yi,Grunberg:2009vs,Grunberg:2011gx,Moch:2009mu,Moch:2009hr,Vogt:2010pe,Soar:2009yh,Vogt:2010cv,Almasy:2010wn}. 

The paper is organized as follows. In section 2 we review briefly the
threshold and joint resummed prompt photon $p_T$ distribution, and
discuss the inclusion of soft-collinear effects.  In section 3 we
assess the numerical impact of these corrections, and we conclude in
section 4. In three appendices we clarify various technical points.

\section{Resummed transverse momentum distributions}
\label{sec:resumm-transv-moment}

We consider the inclusive transverse momentum distribution of prompt photons produced
at fixed $p_T$ in hadron-hadron collisions at center of mass (cm) energy $\sqrt{S}$
\begin{equation}
  \label{eq:el-proc}
  h_A(p_A) + h_B(p_B) \to \gamma(p_c)+X \,,
\end{equation}
where $h_{A,B}$ refers to the two incoming hadrons
 and $X$ to the unobserved part of the final state. 
The lowest order QCD processes producing the prompt photon directly
at partonic cm energy $\sqrt{s}$ are
\begin{equation}
  \label{eq:parton-proc}
  \begin{split}
    &q(p_a) + \bar q(p_b) \to \gamma(p_c) + g(p_d) \\
    &g(p_a) +  q(p_b) \to \gamma(p_c) + q(p_d)\,\,,
  \end{split}
\end{equation}
where in the second reaction $q$ stands for both quark and anti-quark.
The mimimum invariant mass $s$ required for the final state is
$4p_T^2$. It is convenient to express the distance above threshold 
by the variable $1-x_T^2$, where $x_T^2 = 4p_T^2/S$. At the parton level this
becomes $1-\hat{x}_T^2 = 1-4p_T^2/s$.

Apart from the partonic sub processes that directly produce the
photon, there are contributions from $2 \to 2$ parton scattering
\begin{equation}
  \label{eq:frag-proc}
  a(p_a) + b(p_b) \to c(p_c) + d(p_d)\,, 
\end{equation}
where the photon is produced by fragmentation of 
final state parton $c$. In this paper, given the accuracy to which we
work, this will be either a quark or anti-quark. The fragmentation component 
also contributes at $O(\alpha \alpha_s)$, as does the direct component,
though it is subdominant in the sense that the fragmentation function behaves
as $1/N$ \cite{Catani:1999hs}. 
This is in part because, for $pp$ and $pN$ collisions, the fragmentation component 
proceeds via valence quark scattering, as opposed to direct component which involves either
a gluon or a sea quark.
Morever, threshold resummation can substantially 
enhance this fragmentation component \cite{deFlorian:2005wf}.
Here we consider the contribution of the fragmentation
component to the threshold and joint-resummed cross section for prompt
photons at fixed $p_T$ when also the leading soft-collinear effects
are included.  

The resummed cross section consists of two parts
\begin{eqnarray}
  \label{eq:1}
  {p_T^3 d \sigma^{({\rm resum})}_{AB\to \gamma+X} \over dp_T}
  &=&  {p_T^3 d \sigma^{({\rm direct })}_{AB\to \gamma+X} \over dp_T}
  +  {p_T^3 d \sigma^{({\rm frag})}_{AB\to \gamma+X} \over dp_T}\
\end{eqnarray}
where the two terms correspond to the subprocesses \eqref{eq:parton-proc}
and \eqref{eq:frag-proc}, respectively.

The expression for the joint- and threshold-resummed $p_T$ distribution of the direct
component of the prompt photon hadroproduction cross section was 
derived in \cite{Laenen:2000ij} (expressed in a somewhat different
form in \cite{Basu:2007nu}). To be able to compare the expression for
the fragmentation component below in Eq.~\eqref{eq:3} with that for the direct
component,  we quote the latter result again here, but explain
its structure only briefly. To next-to-leading logarithmic accurary, 
with  leading $\ln N/N$ effects included, it reads
\begin{eqnarray}
  \label{eq:23}
    {p_T^3 d \sigma^{({\rm direct})}_{AB\to \gamma + X} \over dp_T}
&=& \frac{p_T^4}{8 \pi S^2}\ \sum_{ab}\  \int_{\cal C} {dN \over 2 \pi i}
\int {d^2 {\bf Q}_T \over (2\pi)^2}\;
\int d^2 {\bf b}\; {\rm e}^{i {\bf b} \cdot {\bf Q}_T} \;
\theta\left(\bar{\mu}-|{\bf Q}_T|\right)
\;\nonumber \\
&\ & \hspace{-25mm} \times\;
\int_0^1 d\tilde x^2_T\; \left(\tilde x^2_T \right)^N
{|M_{ab}(\tilde x^2_T)|^2\over \sqrt{1-\tilde{x}_T^2}}\;
 C_{ab\to \gamma d}(\as(\mu),\tilde
x_T^2)\;\left( \frac{S}{4 |{\bf p}_T - {\bf Q}_T/2|^2} \right)^{N+1}
 \nonumber\\
&\ & \hspace{-25mm} \times
   {\mathcal C}_{a/A}(Q,b,N ) \;  {\mathcal C}_{b/B}(Q,b,N)\; 
\exp\left[E_a^{\rm PT}(N,b,\mu,Q)+E_b^{\rm PT}(N,b,\mu,Q)\right]\nonumber \\[0.5ex]
&\ & \hspace{-25mm} \times
\exp\left[F_d (N,Q,\mu)+ g^{(1)}_{abd} (\lambda)\right]\,.
\end{eqnarray}
Joint resummation resums threshold and recoil logarithms together in
terms of the variable $N$, the Mellin conjugate to $\hat{x}_T^2$, and
the impact parameter $\mathbf{b}$, the Fourier conjugate to $\mathbf{Q_T}$ \cite{Laenen:2000de,Laenen:2000ij}.
The latter is the recoil transverse momentum of the underlying hard scattering process,
over which in the top line the integral is
taken. The variable $\bar{\mu}$ is a cut-off on this transverse
momentum. The hard scale $Q$ is in the present case equal to $2p_T$, and
$\mu$ is the renormalization scale.

The second line contains a Mellin
transform over the partonic scaling variable $\tilde{x}_T^2$ in the recoil
frame (indicated by the tilde), the Born amplitude $M_{ab}$, the $N$- and $b$-independent hard virtual
corrections $C_{ab\to \gamma d}(\as(\mu),\tilde
x_T^2)$ and a kinematic factor linking recoil (through $\mathbf{Q_T}$)
and threshold (through $N$) effects. Finally, the last two lines contain the Sudakov exponentials from
initial and final partons as well as soft wide angle radiation in
combined $(N,b)$ space. They also feature the coefficients ${\mathcal
  C}_{a/A}(Q,b,N ) $ and $ {\mathcal C}_{b/B}(Q,b,N)$ which contain
the evolution matrix for evolution of the parton distribution
functions $ f_{h/H}(N ,\mu_F)$ from scale $\mu_F$ to scale $ Q/\chi$ and the parton
distribution functions. 

The initial state perturbative exponent is
given by
\begin{equation}
\label{eq:31}
      E_a^{\rm PT}(N,b,Q,\mu)  =
\frac{1}{\alpha_s (\mu)}h_a^{(0)} (\beta) +
h_a^{(1)} (\beta,Q,\mu)   \;  ,
\end{equation}
with 
\begin{equation}
\label{eq:38}
\beta = b_0\, \alpha_s (\mu)
\ln \left( \chi \right) \, .
\end{equation}
The functions $h_a^{(0,1)}$  are listed in the Appendix \ref{sec:exponents}. The function $\chi(bQ,N)$ defines the $N$ and $b$
dependent minimum scale of soft gluons to
be included. As in \cite{Basu:2007nu} we choose it to be
\begin{equation}
  \label{eq:6}
\chi(bQ,N)=\bar{b} + \frac{\bar{N}}{1+\eta{\bar{b}}/\bar{N}}\; ,  
\end{equation}
where
\begin{equation}
\label{eq:40}
  \bar{N} = N e^{\gamma_E}, \quad
\bar{b} = bQ e^{\gamma_E}/2\,,
\end{equation}
$\gamma_E$ being the Euler constant. 
The functions $ {\mathcal C}_{a/A}(Q,b,N )$ and $ {\mathcal C}_{b/B}(Q,b,N)$ 
are given by 
\begin{equation}
  \label{eq:5}
         {\mathcal C}_{h/H}(Q,b,N)
=   \sum_{g} {\cal E}_{hg} \left(N,Q/\chi,\mu_F\right) \,
                 f_{g/H}(N ,\mu_F) \; .
\end{equation}
where the matrix $\mathcal{E}$ is the evolution matrix which implements evolution from scale $\mu_F$ to scale $Q/\chi$.
In so doing \cite{Kulesza:2003wn} one includes the leading $\ln N/N$
effects due to initial state radiation in the evolution kernel
\cite{Basu:2007nu}. Note that, as a
consequence, the expressions for $h^{(0)}_a$ and $h^{(1)}_a$ given
here are slightly different from the versions in
Ref.~\cite{Laenen:2000ij}. To the accuracy we work, the $\mu_F$ dependence
essentially cancels in Eq.~\eqref{eq:5}.

The final state exponent reads \cite{Mathews:2004pu}
\begin{equation}
  \label{eq:9}
  F_d (N,Q,\mu) \equiv \frac{1}{\alpha_s(\mu)}f^{(0)}_k(\lambda) + f^{(1)}_k(\lambda,Q,\mu) 
          +f'_k(\lambda,\alpha_s)
\end{equation}
where
\begin{equation}
\label{eq:10}
\lambda = b_0 \alpha_s(\mu^2)\ln \bar{N} 
\end{equation} 
The functions $f_k^{(0)}(\lambda)$ , $f_k^{(1)}(\lambda)$ and
$f'_k(\lambda)$, as well as $g^{(1)}_{abd} (\lambda)$ in
Eq.~\eqref{eq:23}, the exponent due to wide angle soft
radiation, are all listed in Appendix \ref{sec:exponents}.

If one wishes to sum only threshold-enhanced logarithms, an easy
modification of Eq.~\eqref{eq:23} \cite{Basu:2007nu, Laenen:2000ij}
suffices. One merely neglects the recoil term
$\mathbf{Q_T}$ in the kinematic factor in the third line of
Eq.~\eqref{eq:23}, upon which the $\mathbf{Q_T}$ integral sets $\mathbf{b}$
to zero. 

Having reviewed the resummed cross section for the direct component of
prompt photon production let us now turn to the fragmentation component.
The joint-resummed expression  for the fragmentation component was 
derived in Refs.~\cite{Laenen:2000ij,deFlorian:2005yj}. It is given by 
\begin{eqnarray}
\label{eq:3}
 {p_T^3 d \sigma^{({\rm frag})}_{AB\to \gamma+X} \over dp_T}
&=& \sum_{abc} \frac{p_T^4}{8 \pi S^2} \int_{\cal C} {dN \over 2 \pi i}\;
f_{a/A}(N,\mu_F) f_{b/B}(N,\mu_F)D_{\gamma/c}(2(N-1)+3,\mu_F)\;\nonumber\\
&\ & \hspace{5mm} \times
\int {d^2 {\bf Q}_T \over (2\pi)^2}\;
\Theta\left(\bar{\mu}-Q_T\right)
\int d^2 {\bf b} \,
{\rm e}^{i {\bf b} \cdot {\bf Q}_T} \,
\left( \frac{S}{4 |{\bf p}_T-\mathbf{Q_T}/2|^2} \right)^{N+1}
\nonumber \\
&& \hspace{10mm} \times
\; \int_0^1 d\tilde x^2_T \left(\tilde x^2_T \right)^N\;
{|M_{ab\to cd}(\tilde x^2_T)|^2\over \sqrt{1-\tilde{x}_T^2}}\,
\;
C_{ab\to cd}(\as(\mu),\tilde
x_T^2)\nonumber\\
&\ & \hspace{15mm} \times
\exp\left[E_{ab \to cd}\left( N,b,\frac{4 p_T^2}{\tilde
x^2_T},\mu_F \right)\right]\, .
\end{eqnarray}
Here $D_{\gamma/c}(2(N-1)+3,\mu_F)$ is the fragmentation function which expresses
the probability of parton $c$ fragmenting into a photon.  
To include the soft-collinear effects we shall now make modifications
similar to those for the direct component. There is however an
important difference associated with the fragmentation function, which
we shall discuss further below.
Soft-collinear effects due to radiation from initial state partons $a$ and $b$,
can again be included by replacing
the parton distribution functions by the functions $ {\mathcal
  C}_{i/A}(Q,b,N)$  and $ {\mathcal C}_{j/B}(Q,b,N)$, and then of course
modifying the associated exponent functions accordingly. To include the
soft-collinear effects due to radiation from parton $d$, we shall modify
the final state exponent as in Eq.~\eqref{eq:9}.
Finally, for parton $c$, associated with the fragmentation function,
one could envisage two approaches. In the first, one can 
include the soft-collinear effects directly via a modified
exponent function, in analogy to the treatment for parton $d$. This is
the approach we shall study numerically in section \ref{sec:results-1}. 
In a second approach one could attempt to include them instead through evolution, in analogy to
partons $a$ and $b$. We discuss this approach, which is 
problematic, further in appendix \ref{sec:inhom-evol}.

The modification required for the first approach is 
derived as follows \cite{Basu:2007nu,Kramer:1996iq,Catani:2003zt}. Starting 
with the usual integral form for the initial state exponent in threshold resummation \cite{Sterman:1986aj,Catani:1996yz}
\begin{equation}
  \label{eq:43}
E_c^{\mathrm{PT}}(N,Q) =  2 \int_0^1 dz \frac{z^{N-1}-1}{1-z} \int_{\mu_F}^{Q(1-z)} \frac{d\mu}{\mu}A(\alpha_s(\mu)
\end{equation}
one replaces, for a final state fragmenting (anti)quark,
\begin{equation}
  \label{eq:44}
  \frac{z^{N-1}-1}{1-z} \longrightarrow   \frac{z^{N-1}-1}{1-z} -z^{N-1}\,.
\end{equation}
As a consequence, carrying out the integral, one finds
\begin{equation}
\label{eq:46}
  E_c^{\mathrm{PT}}(N,Q,\mu,\mu_F) = E_c^{\mathrm{PT,eik}}(N,Q,\mu,\mu_F) + E_c^{\mathrm{PT,SC}}(N,Q)\,,
\end{equation}
where the first term contains the leading and next-to-leading threshold logarithms, and 
the second term the leading soft-collinear effects. The result is
\begin{equation}
\label{eq:41}
  E_c^{\mathrm{PT,eik}}(N,Q,\mu,\mu_F) = \frac{1}{\alpha_s(\mu)} q^{(0)}(\lambda) + q^{(1)}(\lambda) \,,
\end{equation}
where $\lambda = b_0\alpha_s \ln \bar{N}$. The functions $q^{(0,1)}(\lambda)$ are 
defined in appendix \ref{sec:exponents}. For the soft-collinear effects we find
\begin{equation}
  \label{eq:45}
  E_c^{\mathrm{PT,SC}}(N,Q) = -\frac{A^{(1)}}{2\pi b_0}\exp\Big(-\frac{\lambda}{b_0\alpha_s} \Big) \Big[\ln(1-2\lambda) \Big]\,.
\end{equation}

We now arrive at the expression
\begin{align}
\label{eq:33}
 {p_T^3 d \sigma^{({\rm frag})}_{AB\to \gamma+X} \over dp_T}
&=\sum_{abc} \frac{p_T^4}{8 \pi S^2} \int_{\cal C} {dN \over 2 \pi i}\;
D_{\gamma/c}(2(N-1)+3,\mu_F)\;\nonumber\\
& \hspace{10mm} \times
\int {d^2 {\bf Q}_T \over (2\pi)^2}\;
\int d^2 {\bf b} \,
{\rm e}^{i {\bf b} \cdot {\bf Q}_T}
\Theta\left(\bar{\mu}-Q_T\right)
\left( \frac{S}{4 {\bf p}_T'{}^2} \right)^{N+1}\nonumber\\
& \hspace{10mm} \times
{\mathcal C}_{i/A}(Q,b,N ) \;  {\mathcal C}_{j/B}(Q,b,N) \Sigma^{({\rm resum})}_{ab \to cd}\left( N-1,b \right)\, .
\end{align}
Here, $\Sigma^{({\rm resum})}_{ab \to cd}\left( N,b \right)$ is the resummed cross section for the partonic 
process $ab \to cd$ in combined $N,b$ space and reads
\begin{align}
\label{eq:4}
& \hspace{-1cm} \Sigma^{({\rm resum})}_{ab \to cd}\left( N-1,b \right)  \nonumber \\
&  = \exp\left[E^{\rm PT}_a(N,b,Q,\mu)
+E^{\rm PT}_b (N,b,Q,\mu) +E^{\rm PT}_c(N,Q,\mu,\mu_F)+F_d(N,Q,\mu)\right]
\nonumber\\[0.5ex]
& \times \mathrm{Tr}\left\{ H(Q,\mu)  \bar{\mathrm{P}}\exp\left[
    \int_{p_T}^{p_T/N}\frac{d\mu^{\prime}}{\mu^{\prime}}
\Gamma_{S}^{\dagger}(\alpha_s({\mu^{\prime}}))\right]
S\left(\alpha_s \left(\frac{p_T}{N}\right)\right)\nonumber\right.\\
& \left.\times  \mathrm{P}  \exp\left[
  \int_{p_T}^{p_T/N} \frac{d\mu^\prime}{\mu^\prime}
\Gamma_{S}(\alpha_s({\mu^{\prime}}))\right] \right\}\,.
\end{align}
Here $E^{\rm PT}_a(N,b,Q,\mu)$, $E^{\rm PT}_b (N,b,Q,\mu) $ correspond
to Eq.~\eqref{eq:31},  $F_d(N,Q,\mu)$  to Eq.~\eqref{eq:9} and 
$E^{\rm PT}_c(N,Q,\mu,\mu_F)$ to Eq.~\eqref{eq:46}.
The exponents inside the trace are associated with wide
angle soft radiation and its form is discussed below. 

The expression for the resummed exponent differs crucially from
the corresponding expression for the  direct component in that
there are a larger number of color structures that can connect the external
partons in \eqref{eq:frag-proc} as opposed to direct production. This
requires a set of corresponding coefficient functions and soft
anomalous dimension matrices, which may mix under soft emissions. 
These have already been computed for the
threshold-resummation studies of
\cite{deFlorian:2005yj,deFlorian:2005wf}. In appendix
\ref{sec:diag-soft-anom} we recall the derivation of these factors. 
As a well-known result of this mixing of different color structures 
the radiative factor for wide angle soft radiation takes the form of a
matrix in the space of allowed tensors that connect
the color representations of partons $a,b,c,d$ into a singlet. 
The trace in Eq.~\eqref{eq:4} is taken in that color tensor space.  $S$ is a soft
gluon function that represents non-collinear soft gluon emission, while
$H$ is the hard scattering function describing the short distance hard
scattering.  Both $H$ and $S$ are matrices in color tensor space. At lowest
order, $S_{LI} = \mathrm{Tr}[c_L^\dagger c_I] $ where $c_L^\dagger$ and $ c_I$
are color tensors \cite{Kidonakis:1998nf}. The soft anomalous
dimension matrix $\Gamma_S$ represents the evolution of the soft
function from scale $p_T/N$ to $p_T$. The symbols $\mathrm{P}$ and $\mathrm{\bar{P}}$
denote path ordering in the space of color tensors.
 
Clearly, in a color basis in which the soft anomalous dimension $\Gamma_S$ is
diagonal, the path-ordered exponentials of matrices in
Eq.~\eqref{eq:4} reduce to a sum of simple exponentials. 
Expressions for these soft anomalous dimension matrices have been given in
Ref.~\cite{Kidonakis:1998nf} in terms of the mandelstam invariants
$s,t,u$ associated with the $2\rightarrow 2$ kinematics of reactions
\eqref{eq:parton-proc} and \eqref{eq:frag-proc}. In the threshold limit
$\hat{x}_T^2 \rightarrow 1$ one can approximate these invariants in 
the soft anomalous dimension matrices by 
\begin{equation}
  \label{eq:2}
  s \rightarrow 4p_T^2, \qquad  t \rightarrow -2 p_T^2 , \qquad  u \rightarrow -2 p_T^2 \,.
\end{equation}
We will illustrate this diagonalization procedure for the case of
$qq\rightarrow qq$ in Appendix \ref{sec:diag-soft-anom}.  After the diagonalization
procedure is carried out, the resummed exponent for a given partonic
channel is given by \cite{deFlorian:2005wf}
\begin{align}
 \label{eq:11}
 & \Sigma^{({\rm resum})}_{ab \to cd}\left( N-1,b \right)  = C_{ab \to
   cd} \exp\Bigg[ E^{\rm PT}_a(N,b,Q,\mu)+E^{\rm PT}_b (N,b,Q,\mu)\nonumber \\
&\hspace{5mm}  +E^{\rm PT}_c(N,Q,\mu,\mu_F)+F_d(N,Q,\mu)\Bigg]
\nonumber \\ & \times \left[\sum_I G^I_{ab \to cd}
  \exp\left(\Gamma^{I,(int)}_{ab \to cd}(N) \right) \right]  \sigma^{({\rm Born})}_{ab \to cd}\left( N-1,b \right)
\end{align}
The sum runs over all possible color configurations $I$ with $G^I_{ab \to cd} $ representing a weight for 
each color configuration such that $\sum_I G^I_{ab \to cd} = 1$.
The anomalous dimensions $\Gamma^{I,(int)}_{ab \to cd}(N)$  are given by 
\begin{equation} 
\Gamma^{I,(int)}_{ab \to cd}(N)= \int_0^1 \frac{z^{N-1}-1}{1-z} D_{I,ab \to cd}\left(\alpha_s\left (\left(1-z\right)^2Q^2\right)\right)
\end{equation}
The NLL expansion of $\Gamma^{I,(int)}_{ab \to cd}(N)$ is given by 
\begin{equation}
\Gamma^{I,(int)}_{ab \to cd}(N) = \frac{D_{I,ab \to cd}^{(1)}}{2\pi b_0}\ln(1-2\lambda) + O\left(\alpha_s(\alpha_s \ln N)^k \right)
\end{equation}
The coefficients $D_{ab \to cd}^{(1)}$ , the color weights $G^I$ , one loop hard coefficients $C^{(1)}_{ab \to cd}$ and Born cross sections in $N$- space 
$\Sigma^{({\rm Born})}_{ab \to cd}$ have been given in the Appendix of
Ref.~\cite{deFlorian:2005yj}, for each of the partonic subprocesses in \eqref{eq:frag-proc}.
Eqs.~(\ref{eq:33}) and (\ref{eq:11}) are the expressions we used for
the results in section \ref{sec:results-1}.

Having presented the expressions for joint and threshold resummed production
of prompt photons including soft-collinear effects, for both direct and fragmentation
component,  we next examine the latter numerically.

\section{Numerical studies}
\label{sec:results-1}

In this section we study numerically the inclusion of the $\ln N/N$ terms for the
case of prompt photon production for two kinematic conditions:
those of $p\bar{p}$ collisions at the Tevatron at $\sqrt{S} = 1.96$ TeV
\cite{Abazov:2005wc,Acosta:2004bg},
and those of the $pN$ collisions in the E706 \cite{Apanasevich:2004dr}
fixed target experiment with $E_{\mathrm{beam}} = 530$ GeV,
corresponding to $\sqrt{S} = 31.5$ GeV.
As in our previous study \cite{Basu:2007nu}, we stress that
our aim is primarily to assess the effect of such terms 
in relevant kinematic conditions, rather than perform a comprehensive comparison with data.
Our assessments therefore mainly consist of comparing the same calculation 
with and without $\ln N/N$ terms, and compare this with the difference
between LL and NLL accuracy.

Our default choices for various input parameters are as follows. 
We use joint resummation unless specified otherwise.
We use the GRV parton density set \cite{Gluck:1998xa}, corresponding to
$\alpha_s(M_Z)=0.114$, with the evolution code of
Ref.~\cite{Vogt:2004ns}, changing flavor number at $\mu = m_c\; (1.4
\,\mathrm{GeV})$ and $m_b \;(4.5 \, \mathrm{GeV})$.  We choose the
factorization and renormalization scale equal to $p_T$, and 
the non-perturbative parameter $g_{\mathrm{NP}}$ in
Eq.~\eqref{eq:37} equal to $1\,\mathrm{GeV}^2$. 
For the parameter $\chi$ we use 
the expression in Eq.~\eqref{eq:18}, following \cite{Kulesza:2002rh},
with $\eta = 1/4$.
For our joint-resummed results, we chose for Tevatron (E706) kinematics 
the cut-off $\bar{\mu}$ in Eq.~\eqref{eq:6} equal to $15\, (5)$ GeV. 
Regarding logarithmic accuracy, and unless stated otherwise, we refer to LL when
using only $h^{(0)}_a$,  $f^{(0)}_k$  and $q^{(0)}$, and
$\bar{C}^{(ab\to \gamma d)} = \bar{C}^{(ab\to cd)} = 1$ (see Appendix
\ref{sec:exponents}); 
we refer to NLL  when also including $h^{(1)}_a$, $f^{(1)}_k$, $q^{(1)}$ and the
virtual corrections discussed below Eq.~\eqref{eq:32}.
When including the soft-collinear $\ln N/N$ terms we use the
full NLO anomalous dimension, evolved from scale $\mu_F$ to $Q/\chi$ in Eq.~\eqref{eq:16}. 

\begin{figure}
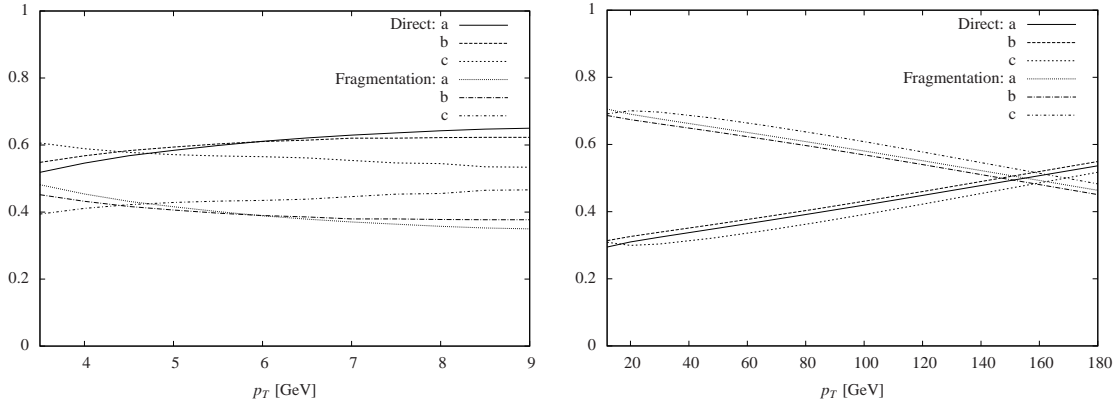

  \centering
  \includegraphics[width=0.452\textwidth]{Fig1a.epsi}
  \hspace{0.25cm}
  \includegraphics[width=0.46\textwidth]{Fig1b.epsi}
  \caption{Relative contributions vs. $p_T$ of direct and
    fragmentation photons for LL (a), NLL (b), NLL + $\ln N/N$ (c). 
The left pane shows the results for E706 kinematics and the
    right pane for Tevatron kinematics. }  
\label{fig:fig1}
\end{figure}

In Fig.~\ref{fig:fig1} we show the relative contributions of the
direct and fragmentation components for both kinematic conditions to
the total result
\begin{equation}
  \label{eq:27}
  \frac{\mathrm{direct}}{\mathrm{direct\,+\,fragmentation}}, \quad 
  \frac{\mathrm{fragmentation}}{\mathrm{direct\,+\,fragmentation}}\,,
\end{equation}
for LL, NLL, as well as NLL with soft-collinear effects.
For E706 kinematics we observe that, as $p_T$ increases, the contribution
from the direct component becomes more dominant, but not such that
the fragmentation component becomes wholly negligible. We note in
particular that in the case of NLL with $\ln N/N$ the relative direct contribution decreases slightly over the range of $p_T$ values.\\
In the case of Tevatron kinematics,  for values of $p_T
\leq$ 150 GeV the relative direct contibution is larger than the
relative fragmentation contribution. The contribution of
soft-collinear effects is mostly larger than the difference between LL
and NLL, the latter difference being quite small for joint resummation.
\begin{figure}
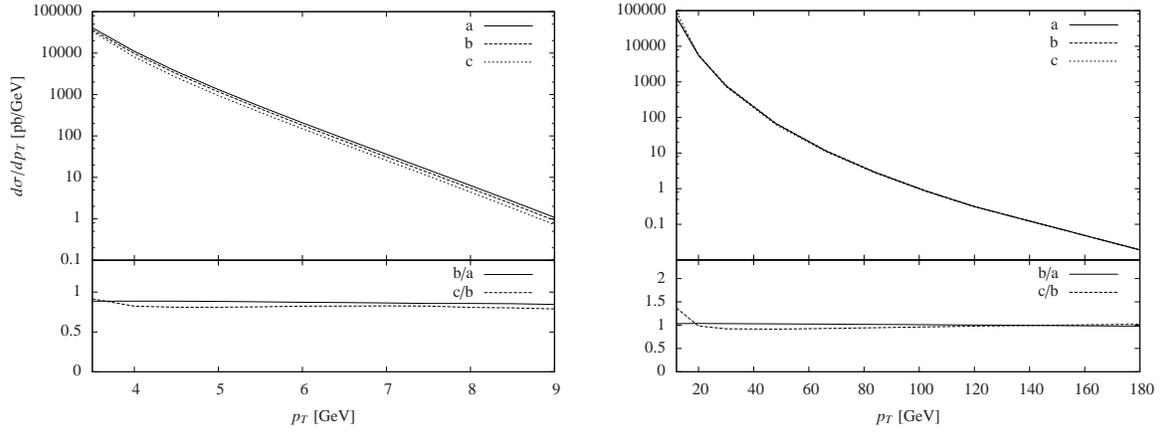

  \centering
  \includegraphics[width=0.47\textwidth]{Fig2a.epsi}
  \hspace{0.5cm}
  \includegraphics[width=0.46\textwidth]{Fig2b.epsi}
  \caption{The LL (a), NLL (b) and NLL+$\ln N/N$ (c) calculations as well as the ratios NLL/LL, NLL+$\ln N/
    N$/NLL vs. $p_T$. Left pane: E706, right pane: Tevatron.}
  \label{fig:fig2}
\end{figure}
In Fig.~\ref{fig:fig2} we show the fragmentation component by itself
for both kinematics, and for our three approximations.
In the case of E706 kinematics we note that the NLL curve is decreased slightly relative to the LL curve. Furthermore, inclusion of the soft-collinear contribution leads to a further small decrease.
In the case of Tevatron kinematics we see that the difference between
the curves is even smaller, due to the smaller value of $\alpha_s$ at larger
$p_T$ values. 
Inclusion of the $\ln N/N$ terms lowers the result with respect to the
NLL curve for $p_T$ above 20 GeV by a few percent,
but their contribution is negligible for $p_T$ values above 110 GeV. 
\begin{figure}
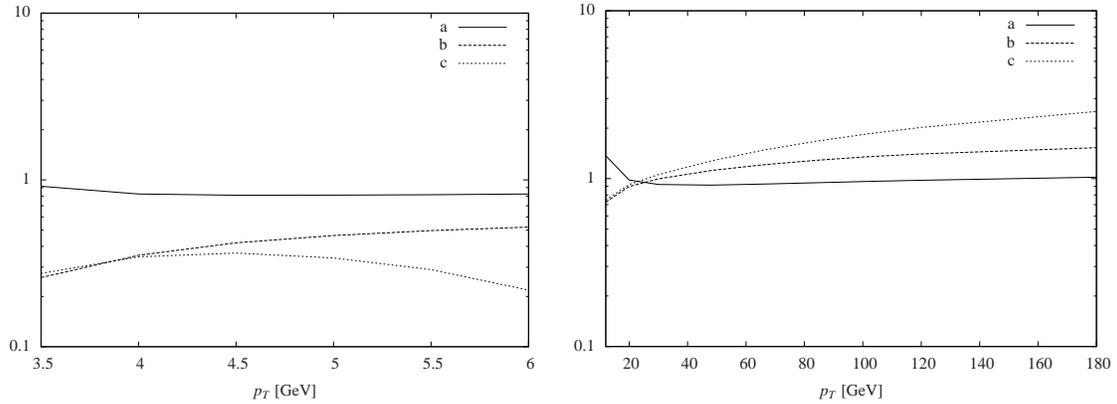

  \centering
  \includegraphics[width=0.45\textwidth]{Fig3a.epsi}
  \hspace{0.25cm}
  \includegraphics[width=0.46\textwidth]{Fig3b.epsi}
  \caption{Comparison of joint resummation and threshold resummation effects, ratios to joint resummation NLL without $\ln N/N$. The three curves show joint NLL with $\ln N/N$ (a), threshold NLL without $\ln N/N$ (b) and threshold NLL with $\ln N/N$ (c). Left pane: E706, right pane: Tevatron.}
  \label{fig:fig3}
\end{figure}
In Fig.~\ref{fig:fig3} we show a comparison between joint- and
threshold resummation for both E706 and Tevatron kinematics. The curves are all relative to the joint NLL result without soft-collinear terms.
While for E706 and joint resummation the inclusion of the $\ln N/N$ terms leads to only a small
decrease of the NLL curve , we see that the
effects are much more appreciable for threshold resummation.
In particular we note that for $p_T$ above 4 GeV the inclusion
of $\ln N/N$ terms leads to a notable decrease of the NLL curve. This is
very similar to what we found for the direct component in Ref.\cite{Basu:2007nu}.
For Tevatron kinematics the inclusion of $\ln N/N$ terms for joint
resummation is only appreciable for quite low values of $p_T$.
For threshold resummation the situation is again notably different,
with the inclusion of the soft-collinear terms leading to an
appreciable increase relative to the pure NLL result, again in
similarity to the direct component \cite{Basu:2007nu}.
\begin{figure}
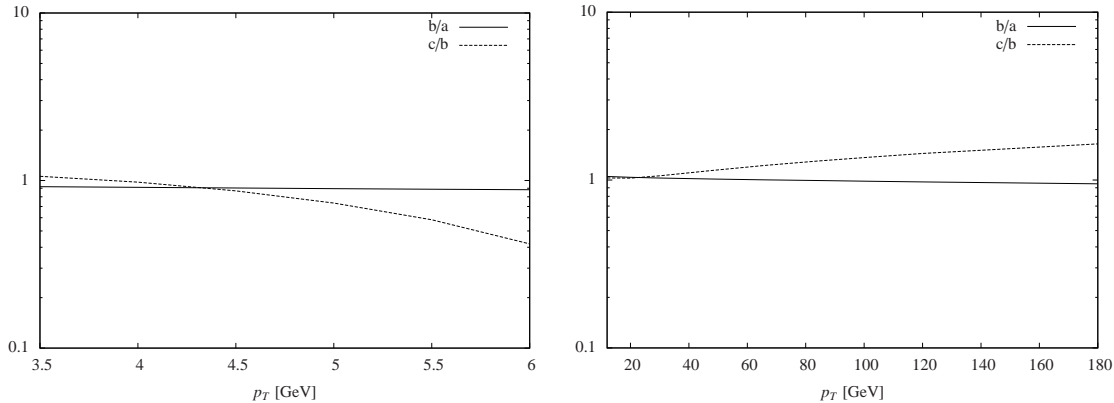

  \centering
  \includegraphics[width=0.452\textwidth]{Fig4a.epsi}
  \hspace{0.25cm}
  \includegraphics[width=0.46\textwidth]{Fig4b.epsi}
  \caption{Same as the ratio plots in Fig.~\ref{fig:fig2} but for threshold resummation. Left pane: E706, right pane: Tevatron.}
  \label{fig:fig4}
\end{figure}
Finally, Fig.~\ref{fig:fig4} is equivalent to the ratios shown in
Fig.~\ref{fig:fig2} but now for threshold resummation only.
The soft-collinear terms change from increasing to decreasing the
$p_T$ distribution for E706 kinematics, with the opposite pattern
occurring for Tevatron kinematics.

The behavior in Figs.~\ref{fig:fig3} and~\ref{fig:fig4}
is quite similar to the direct case in our previous paper, which is perhaps not surprising since
the resummation effects  (LL, NLL and $\ln N/N$) are
introduced by means of process independent 
exponentials and evolution of parton distribution functions. 
The extra contributions from the fragmentation parton and the
additional color structures do not seem to change these patterns noticeably.
 
\section{Conclusions}
\label{sec:conclusions}

We have examined the effects of including terms of the form
\begin{equation}
\label{eq:26}
  \alpha_s^i \sum_j^{2i-1}\, d_{ij} \frac{\ln^jN}{N}\,.
\end{equation}
in the production of a hard photon through fragmentation for 
$p_T$ distributions at both collider and fixed target kinematics,
at leading accuracy ($j=i$), both for joint and threshold resummation.
This is the complement to an earlier
study carried for the direct production component \cite{Basu:2007nu}.

To the extent that leading terms of the form \eqref{eq:26} arise from initial state radiation
effects, we used the method of Refs.~\cite{Kulesza:2002rh,Kulesza:2003wn}
to include them. Those arising from final state emission we included
both by extending the resummation exponents for the fragmenting parton
and the unobserved parton to leading $\ln N/N$ accuracy. 
As for the direct component, we found the combined $\ln N/N$ terms to be 
comparable to NLL corrections, and dependent on kinematics, either 
enhancing or suppressing. The contribution from $\ln N/N$ terms is
noticable in the case of joint resummation although it remains small. This seems to imply that the corrections introduced by recoil effects tend to overshadow those of the soft-collinear terms.
In the case of pure threshold resummation the soft-collinear effects  play a  more appreciable role.
Overall we found a behavior comparable to the direct case studied previously.

\subsection*{Acknowledgments}

We thank Daniel de Florian and Werner Vogelsang for helpful
discussions.  AM would like to thank Department of Atomic Energy-BRNS, India
for financial support under the grant No. 2010/37P/47/BRNS. We would like
to thank the Institute for Mathematical Sciences in Chennai for
gracious hospitality. PM is grateful to the department of physics at Mumbai
University for support during a visit. EL has been supported by the 
National Organization for Scientific Research (NWO), 
and the Foundation for Fundamental Research 
of Matter (FOM), program 104 ``Theoretical Particle Physics in the Era
of the LHC''. PM has been supported by the Bundesministerium f\"{u}r Bildung und Forschung (BMBF).

\appendix

\section{Exponents}
\label{sec:exponents}

Here we list the exponents used in section \ref{sec:resumm-transv-moment}.
The initial state exponents for the LL and NLL case without inclusion of the $\ln N/N$--terms are given by
\begin{eqnarray}
\label{eq:14a1}
h_a^{(0)} (\lambda,\beta) & =& \frac{A_a^{(1)}}{2\pi b_0^2}\Big[2\beta + (1-2\lambda)\ln(1-2\beta) \Big] \\
\label{eq:14a2}
h_a^{(1)} (\lambda,\beta,Q,\mu,\mu_F)& =& \frac{1}{2\pi
  b_0}\left(-\frac{A^{(2)}_a}{\pi b_0} + A^{(1)}_a \ln\Big(\frac{Q^2}{\mu^2} \Big) \right)\Big[\frac{2 \beta (1-2 \lambda)}{(1-2 \beta)} + \ln(1-2\beta) \Big] \nonumber \\
            &  +& \frac{A^{(1)}_a b_1}{2\pi b_0^3}\Big[\frac{(1-2 \lambda)(2\beta + \ln(1-2\beta))}{(1-2 \beta)}+\frac{1}{2}\ln^2(1-2\beta) \Big] \nonumber \\
            &  -& \frac{A^{(1)}_a}{\pi b_0}\lambda \ln\Big(\frac{Q^2}{\mu_F^2} \Big) \; ,
\end{eqnarray}
while for the case when evolving the parton distribution functions down to $Q/\chi$ (Eq.~\eqref{eq:31}) they are
\begin{eqnarray}
\label{eq:14b}
h_a^{(0)} (\beta) &=& \frac{A_a^{(1)}}{2\pi b_0^2}
\left[ 2 \beta + \ln(1-2 \beta) \right]\, ,\\
h_a^{(1)} (\beta,Q,\mu) &=&
\frac{A_a^{(1)} b_1}{2\pi b_0^3} \left[ \frac{1}{2} \ln^2 (1-2 \beta) +
\frac{2 \beta + \ln(1-2 \beta)}{1-2\beta} \right] + 
\frac{B_a^{(1)}}{2\pi b_0}  \ln(1-2 \beta) \nonumber \\
&+& \frac{1}{2\pi b_0} \left[ A_a^{(1)}\ln \left( \frac{Q^2}{\mu^2} \right)
-\frac{A_a^{(2)}}{\pi b_0}\right] \;
\left[ \frac{2 \beta}{1-2\beta}+ \ln(1-2 \beta) \right] \; .
\end{eqnarray}
where $\beta = b_0 \alpha_s(\mu) \ln \chi$, and 
\begin{equation}
\label{eq:37}
    A_a^{(1)} = C_a, \qquad
    A_a^{(2)} = 
\tfrac{1}{2}C_a \left[C_A\Bigg(\frac{67}{18}-\frac{\pi^2}{6}\Bigg)-\frac{10}{9}T_R N_F \right]
\end{equation}
with $C_q = C_F$ and $C_g = C_A$. Also we have
\begin{equation}
  \label{eq:20}
  B_q^{(1)} = -\frac{3}{4}C_F, \qquad
  B_g^{(1)} = -\pi b_0 \,.
\end{equation}
In these equations
\begin{eqnarray}
b_0 &=& \frac{11 C_A - 4 T_R N_F}{12 \pi}\;\;\;\; , \;\;\;\;\;
b_1 \;=\; \frac{17 C_A^2-10 C_A T_R N_F-6 C_F T_R N_F}{24 \pi^2}\; .
\end{eqnarray}
where $T_R = 1/2$.\\
The functions $q^{(0,1)}(\lambda)$ in Eq.~\eqref{eq:41} are obtained
by setting $\beta=\lambda$ in Eq.~\eqref{eq:14a1} and~\eqref{eq:14a2}.\\
The final state exponents \eqref{eq:9} involve the functions
\begin{equation}
\label{eq:39}
f^{(0)}_a= -\frac{A^{(1)}_a}{2\pi b_0}
          [(1-2 \lambda)\ln(1-2 \lambda) -2(1- \lambda)\ln(1- \lambda)]
\end{equation}   
\begin{eqnarray}
\label{eq:7}
f^{(1)}_a =& -&\frac{A^{(1)}_a b_1}{2\pi b_0^3 }
            [\ln(1-2 \lambda) -2\ln(1- \lambda)
         +\frac{1}{2}\ln^2(1-2 \lambda)-\ln^2(1-\lambda)] \nonumber \\
        &+&\frac{B^{(1)}_a}{ 2 \pi b_0 }\ln(1-\lambda)
      -\frac{A^{(2)}_a}{2\pi^2 b_0^2}[2 \ln(1-\lambda) - \ln(1-2\lambda)] \nonumber \\
&+&\frac{A^{(1)}_a}{2\pi b_0}[2\ln(1- \lambda)- \ln(1-2\lambda)]\ln\frac{Q^2}{\mu^2}
\end{eqnarray}
with $\lambda = b_0 \alpha_s(\mu) \ln \bar{N}$. Note that $\beta$ only 
differs from $\lambda$ for joint resummation; for threshold
resummation $\beta = \lambda$. The final term in
Eq.~\eqref{eq:9} reads
\begin{equation}
\label{eq:25}
f^\prime_q 
       =\frac{A^{(1)}_q}{ 2 \pi b_0}
  \exp\left(-\frac{\lambda}{\alpha_s b_0}\right)\left[\ln(1-2 \lambda)-\ln(1- \lambda)\right]\,,
\end{equation}
\begin{equation}
\label{eq:42}
f^\prime_g 
       =\frac{3 A^{(1)}_g}{ 2 \pi b_0}
  \exp\left(-\frac{\lambda}{\alpha_s b_0}\right)\left[\ln(1-2 \lambda)-\ln(1- \lambda)\right]\,.
\end{equation}
The wide-angle soft radiation exponents in Eq.~\eqref{eq:23}  are
\begin{equation}
  \label{eq:8}
g^{(1)}_{q\bar{q}g}(\lambda) = -\frac{C_A}{\pi b_0}\ln (1-2\lambda) \ln 2,\qquad
g^{(1)}_{qgq}(\lambda) = -\frac{C_F}{\pi b_0}\ln (1-2\lambda) \ln 2
\end{equation}
These expressions are obtained by expanding the perturbative functions 
$A_a(\alpha_s)$, $B_d(\alpha_s)$ and $D_{ab\rightarrow d \gamma}$ in powers of 
$\alpha_s$ 
\begin{equation}
\label{eq:32}
A_a(\alpha_s) = \frac{\alpha_s}{\pi}A^{(1)}_a + 
              \left(\frac{\alpha_s}{\pi}\right)^2 A^{(2)}_a +  O(\alpha^3_s)\,.
\end{equation}
The explicit forms of $C^{(ab\to \gamma d)}$ are shown in \cite{Catani:1998tm,Catani:1999hs}. The $C^{(ab\to cd)}$ are calculated by expanding the resummed cross section (Eq.~\eqref{eq:33}) to $\mathcal{O}(\alpha_s^3)$ and matching to the fixed order NLO result in \cite{Aversa:1988vb}.
We note that there is no factorization scale dependence in $h_a^{(1)}$ and the coefficient
functions in Eq.~\eqref{eq:23} because of complete evolution from scale $\mu_F$ to $Q/\chi$
in Eqs.~\eqref{eq:10},\eqref{eq:2} and \eqref{eq:5}.

\section{Diagonization of soft anomalous dimensions}
\label{sec:diag-soft-anom}

In this appendix we discuss the diagonalization procedure for the
anomalous dimension matrices of Ref.~\cite{Kidonakis:1999hq}
to arrive at the expressions of 
$G^I_{ab\to cd}$ and $D_{ab \to cd}^{(1)}$ of Ref.~\cite{deFlorian:2005yj}, which we
also use here. In what follows, we illustrate this procedure for
the partonic sub process $qq \rightarrow qq$ in the limit 
\eqref{eq:2}.

When one changes to a new color basis in which $\Gamma_S$ is diagonal the
resummed cross section takes a simpler form.  The diagonalization
procedure has been discussed in detail in Ref.~\cite{Kidonakis:1999hq}.
Upon diagonalization the trace in Eq.~(\ref{eq:4}) reduces to a simple sum of exponentials as in
Eq.~(\ref{eq:11}).  The soft anomalous dimension matrix for the process
$qq\rightarrow qq$ expressed in the $t$-channel singlet-octet basis is given by 
 \begin{equation}
      \label{eq:gamfinqqbar}
 \Gamma_{S'}=\frac{\alpha_s}{\pi}\left(
                 \begin{array}{cc}
                  2C_FU-\frac{1}{N_c}(T+U) &   2 U  \\
                 \frac{C_F}{N_c}U    & 2C_FT\\
                 \end{array} \right)
 \end{equation}
where
\begin{equation}
  \label{eq:12}
U = \ln\left(-\frac{u}{s}\right) + i \pi  , \qquad T = \ln\left(-\frac{t}{s}\right) + i \pi\,.
\end{equation}

This results in a slight difference from the initial state exponent in
Ref.~\cite{deFlorian:2005wf} which we are using. The difference for
the partonic sub process $ab \rightarrow cd $ is then given by 
\begin{equation}
  \label{eq:13}
\frac{\ln 2}{2}\left(C_a + C_b - C_c+ C_d \right)  \ln\left(1-2\lambda\right)  
\end{equation}
which we add to the diagonal terms of
anomalous dimension matrix. This leads to
\begin{equation}
\label{eq:15}
\Gamma_{S'}=\frac{\alpha_s}{\pi}\left(
                \begin{array}{cc}
                 -\frac{2}{3} &   -2 \\
                -\frac{4}{9}  &-\frac{4}{3}
                \end{array} \right)\, \ln 2.
\end{equation}
The eigenvalues of this matrix \eqref{eq:15} are $\lambda_1 = 0$  and
$\lambda_2 = -2$, and the  corresponding eigen vectors are, respectively
\begin{equation}
  \label{eq:16}
 \left(\begin{array}{cc}
                  -3\\
                  1
                \end{array} \right), \qquad \left( \begin{array}{cc}
                 \frac{3}{2} \\
                  1
                \end{array} \right)\,.
\end{equation}
Now,  we change the basis  from $t$-channel singlet-octet basis to 
a new basis in which $\Gamma_S $ is diagonal by using the matrix of eigenvectors 
\begin{equation}
\label{eq:17}
{\cal R}^{-1} = \left(
                \begin{array}{cc}
                 -3  &   -\frac{3}{2} \\
                  1    &    1
                \end{array} \right)\, .
    \end{equation}
Thus we obtain
\begin{align}
 & \mathrm{Tr}\left\{ H(p_T,\mu) \bar{\mathrm{P}}\exp\left[
    \int_{p_T}^{p_T/N}\frac{d\mu^{\prime}}{\mu^\prime}{\Gamma_{S}}^{\prime
      \dagger}
\left(\alpha_s\left({\mu^{\prime}}\right)\right)\right]\right. \nonumber\\
& \hspace{1cm} \times \left.
S\left(\alpha_s \left(\frac{p_T}{N}\right)\right) \mathrm{P}  \exp\left[ \int_{p_T}^{p_T/N} \frac{d\mu^\prime}{\mu^\prime}\Gamma_{S^{\prime}}\left(\alpha_s\left({\mu^{\prime}}^ 2\right)\right)\right] \right\}
\nonumber\\
& = \mathrm{Tr}\left\{
{\cal R}^{-1} H(p_T,\mu) {\cal R}  {\cal R}^{-1}\exp\left[ \int_{p_T}^{p_T/N}\frac{d\mu^\prime}{\mu^\prime}\Gamma_{S}^{ \prime \dagger}\right]{\cal R}
{\cal R}^{-1} S  \exp\left[ \int_{p_T}^{p_TN}
  \frac{d\mu^\prime}{\mu^\prime}\Gamma_{S}^{\prime d}\right]{\cal R}
\right\}
\end{align}
For the present case, 
\begin{equation}
\Gamma_S^{\prime d} = {\cal R}^{-1} {\Gamma_S}^{\prime} {\cal R} =  \left(
                \begin{array}{cc}
                   0  &      0 \\
                   0  &      2
                \end{array} \right)\, 
\end{equation}
$H$ and $S$ are the Mellin moments of $2 \times 2$ hard and soft
matrices in color tensor space\cite{Kidonakis:1999hq}. 
Substituting for them, one finds that the trace finally reduces to a sum of exponents
\begin{align}
\hat \sigma_1\exp\left[\frac{-\lambda_1\ln 2}{2\pi b_0}\ln(1-2\lambda)\right] +\hat \sigma_2\exp\left[\frac{-\lambda_2\ln 2}{2\pi b_0}\ln(1-2\lambda)\right] 
\end{align}
where $\hat \sigma_1$ and $\hat \sigma_2$ are $H'_{1i}S'_{i1}$ and $H'_{2i}S'_{i2}$ respectively, $H'$  and $S'$  being the hard and soft matrices in the new basis. 
Identifying $\frac{\hat \sigma_1}{\sigma_{born}}$ and $\frac{\hat \sigma_2}{\sigma_{born}}$ with $G^1_{qq \rightarrow qq} $ and $G^2_{qq \rightarrow qq} $ and 
$\frac{-\lambda_1\ln 2}{2\pi b_0}$ and $\frac{-\lambda_2\ln 2}{2\pi b_0}$  with $D^1_{qq \rightarrow qq} $ and $D^2_{qq \rightarrow qq} $,  one  obtains the relevant term in the exponent in Eq.~\eqref{eq:11}.

\section{Soft-collinear effects in photon fragmentation function}
\label{sec:inhom-evol}

In this appendix we discuss the possibility of
including the leading soft-collinear effects in a manner exactly
analogous to the initial state in \cite{Basu:2007nu}, i.e. 
$Q/\bar{N}$. The present case is however  special
because the evolution equation for the photon fragmentation function 
is inhomogeneous. 

To leading order, the non-singlet evolution equation for the photon fragmentation function
$D_{\gamma/c}(N,\mu_F)$, where $c$ is the parton that fragments into a
photon, reads,  in moment space
\begin{equation}
  \label{eq:19}
  \frac{d D_{\gamma/c}(N,\mu^2)}{d\ln \mu^2} = \frac{\alpha}{2\pi}
  k^{(0)}(N) + \frac{\alpha_s(\mu^2)}{2\pi} P^{(0)}(N) D_{\gamma/c}(N,\mu^2)
\end{equation}
where to $\mathcal{O}(1/N)$, for 3 active flavors
\begin{equation}
  \label{eq:18}
   k^{(0)}(N) = \frac{4}{3N}, \qquad   P^{(0)}(N) = C_F \left(-2\ln
     \bar{N} - \frac{1}{N} + \frac{3}{2} \right)\,.
 \end{equation}
 The inhomogeneous term arises because the photon, in contrast to
 hadrons, has a pointlike interaction with a quark.
Correspondingly, the solution to Eq.~\eqref{eq:19} is the sum of a homogeneous (hadronic) and inhomogeneous
(pointlike) part
\begin{equation}
  \label{eq:21}
    D^{\gamma}(N,\mu^2) = D^{\gamma}_{\textrm{had}}(N,\mu^2)+D^{\gamma}_{\textrm{pl}}(N,\mu^2)   \,.
\end{equation}
To leading logarithmic accuracy the homogeneous solution is
\begin{equation}
  \label{eq:22}
  D^{\gamma}_{\textrm{had}}(N,\mu^2) = L^{-\frac{P^{(0)}(N)}{2\pi b_0}} D^{\gamma}_{\textrm{had}}(N,\mu_0^2)  
\end{equation}
where
\begin{equation}
  \label{eq:24}
  L = \frac{\alpha_s(\mu^2)}{\alpha_s(\mu_0^2)} = 1 - b_0 \alpha_s(\mu_0^2) \ln \left ( \frac{\mu^2}{\mu^2_0} \right)
\end{equation}
To see what evolution from $Q$ to $Q/\bar{N}$ corresponds to, we substitute
$\mu_0 = Q$ and
$\mu = Q/\bar{N}$  and find
\begin{align}
\label{eq:28}
  L^{-\frac{P^{(0)}(N)}{2\pi b_0}}&  = \left( 1 + 2 b_0 \alpha_s \ln(\bar{N}) \right)^{\frac{1}{\pi b_0}C_F
    \left(\ln(\bar{N})+\frac{1}{2N}-\frac{3}{4} \right)} \nonumber \\
&  \simeq \exp \left(\frac{\alpha_s C_F}{\pi} \left[2 \ln^2(\bar{N}) + \frac{\ln(\bar{N})}{N}-\frac{3}{2}\ln(\bar{N}) \right] \right)
\end{align}
which indeed sums the leading soft-collinear effects. 

We now discuss the point-like (inhomogeneous) solution to
Eq.~\eqref{eq:19}. To LL accuracy it reads
\begin{eqnarray}
  \label{eq:pl1}
  D^{\gamma}_{\textrm{pl}}(N,\mu^2) = \frac{4 \pi}{\alpha_s} \left[ 1
    - L^{1 - \frac{P^{(0)}(N) }{2\pi b_0}} \right] \frac{1}{1-\frac{P^{(0)}(N)}{2\pi b_0}} \frac{\alpha}{8\pi^2 b_0} k^{(0)}(N)\,.
\end{eqnarray}
Notice that $D^{\gamma}_{\textrm{pl}}(N,\mu_0^2) = 0$, and that it is
indeed proportional to $1/N$ via $k^{(0)}(N)$, see Eq.~\eqref{eq:18}.
For $\mu_0 = Q$ and $\mu=Q/\bar{N}$ the factor in square brackets can be rewritten as follows
\begin{eqnarray}
  \label{eq:pl2}
\left[ 1 - L^{1 - \frac{1}{2\pi b_0}P^{(0)}(N)} \right] =  \left[ L^{
    \frac{P^{(0)}(N)}{2\pi b_0}} - \left(1 + 2 b_0 \alpha_s \ln(\bar{N}) \right) \right] L^{ -\frac{P^{(0)}(N)}{2\pi b_0}}\,.
\end{eqnarray}
We can rewrite the contents of square brackets as
\begin{align}
  \label{eq:pl3}
&   1 - \frac{ \alpha_s  C_F}{\pi} \left[2 \ln^2(\bar{N}) +
    \frac{\ln(\bar{N})}{N}-\frac{3}{2}\ln(\bar{N}) \right ] - \left(1
    + 2 b_0 \alpha_s \ln(\bar{N}) \right) \nonumber \\
 &= -\frac{\alpha_s}{\pi} \left[  2C_F \ln^2(\bar{N}) + C_F
   \frac{\ln(\bar{N})}{N}-\frac{3}{2}C_F\ln(\bar{N})  + 2\pi b_0
   \ln(\bar{N})  \right]\, .
\end{align}
The expression in Eq.~(\ref{eq:pl1}) then becomes
\begin{align}
  \label{eq:pl4}
&  D^{\gamma}_{\textrm{pl}}\left(N,\frac{Q^2}{\bar{N}^2}\right) =
\nonumber \\
&   \left( \frac{-4 \left[  2C_F \ln^2(\bar{N}) +C_F
        \frac{\ln(\bar{N})}{N}-\frac{3}{2}C_F \ln(\bar{N}) + 2\pi b_0 \ln(\bar{N}) \right] }{1+\frac{C_F}{\pi b_0}\left ( \ln(\bar{N}) + \frac{1}{2N}-\frac{3}{4} \right ) } \frac{\alpha}{8\pi^2 b_0} k^{(0)}(N) \right) L^{-\frac{P^{(0)}(N)}{2\pi b_0}}
\end{align}
Carrying out the division inside the brackets for large $N$  yields
\begin{eqnarray}
  \label{eq:pl5}
  D^{\gamma}_{\textrm{pl}}(N) = -\frac{\alpha}{\pi} \ln(\bar{N}) k^{(0)}(N) L^{-\frac{P^{(0)}(N)}{2\pi b_0}}\,.
\end{eqnarray}
Taken together the photonfragmentation function takes the form
\begin{eqnarray}
  \label{eq:pl6}
  D^{\gamma}\left(N,\frac{Q^2}{\bar{N}^2}\right) =  D^{\gamma}_{\textrm{had}}\left(N,\frac{Q^2}{\bar{N}^2}\right) -\frac{4}{3}\frac{\alpha}{\pi}\frac{ \ln(\bar{N})}{N} L^{-\frac{P^{(0)}(N)}{2\pi b_0}}\,.
\end{eqnarray}

Let us now perform a rough assessment of the validity of this approach
to include soft-collinear effects.
As the hadronic component of the photon fragmentation function has the property that
one can sum the leading $\ln N/N$ effects through evolution, we 
factor it out of the full solution in \eqref{eq:pl6}, and estimate the large $N$ behavior of the
remainder based on a reasonable assumption of the non-perturbative $N$-dependence.
We parametrize the
hadronic part (stemming from the light quark flavors) in this limit as in
Ref.~\cite{Gluck:1992zx}  by $D_{\textrm{had}}\sim k_1 x^{-0.3} (1 -
x)^2 + k_2 \sqrt{x} (1.703 - x)$. It is now straightforward to show, after a
Mellin transform, that the term for the point-like part contributes
effectively at large $N$ by a multiplicative factor  
\begin{equation}
  \label{eq:14}
1+  \alpha_s (C\ln \bar{N} + \frac{C'}{N}\ln \bar{N})
\end{equation}
to the hadronic component for some constants $C,C'$. This affects the
leading $\ln N/N$ term in \eqref{eq:28}, and would therefore 
seem to make this approach problematic.

\bibliographystyle{h-physrev4}
\bibliography{spires}

\end{document}